\documentclass[proof]{WileyASNA-v1}
\usepackage{bm}
\usepackage[utf8]{inputenc}

\articletype{Proceedings}%

\received{31 December 2020}
\revised{31 December 2020}
\accepted{31 December 2020}

\newcommand{\Blue}[1]{{ #1}}

\begin{document}

%---------------------------------------------------------------------
\title{Precession of triaxially deformed neutron stars
\protect\thanks{Poster presentation at the 9th International Workshop on Astronomy and Relativistic
Astrophysics, 6--12 September, 2020.}}

\author[1,2]{Yong Gao$^{*,}$}
\author[2,3]{Lijing Shao}

\authormark{Y. Gao \& L. Shao}

\address[1]{\orgdiv{Department of Astronomy, School of Physics},
\orgname{Peking University}, \orgaddress{\state{Beijing 100871},
\country{China}}}

\address[2]{\orgdiv{Kavli Institute for Astronomy and Astrophysics},
\orgname{Peking University}, \orgaddress{\state{Beijing 100871},
\country{China}}}

\address[3]{\orgdiv{National Astronomical Observatories}, \orgname{Chinese
Academy of Sciences}, \orgaddress{\state{Beijing 100012}, \country{China}}}

\corres{*\email{gaoyong.physics@pku.edu.cn}}
%---------------------------------------------------------------------
\abstract{A deformed neutron star (NS) will precess if the instantaneous
spin axis and the angular momentum are not aligned. Such a precession can
produce continuous gravitational waves (GWs) and modulate electromagnetic
pulse signals of pulsars. In this contribution we extend our previous
work in a more convenient parameterization. We treat NSs as rigid triaxial
bodies and give analytical solutions for angular velocities and Euler
angles. We summarize the general GW waveforms from freely precessing
triaxial NSs and use Taylor expansions to obtain waveforms with a small
wobble angle. For pulsar signals, we adopt a simple cone model to study the
timing residuals and pulse profile modulations. In reality, the
electromagnetic torque acts on pulsars and affects the precession behavior.
\Blue{Thereof, as an additional extension to our previous work, we consider
a vacuum torque and display an illustrative example for the residuals of
body-frame angular velocities.} Detailed investigations concerning
continuous GWs and modulated pulsar signals from forced precession of
triaixal NSs will be given in future studies.}

\keywords{gravitational waves -- pulsars: general -- methods: analytical}

\jnlcitation{\cname{%
\author{Yong Gao}} (\cyear{2020}), 
\ctitle{Precession of triaxially deformed Neutron stars}, \cjournal{Astron. Nachrichten},
\cvol{2020;ZZZ:XX--YY}.}

%%\fundingInfo{Funding info text.}

\maketitle

\footnotetext{\textbf{Abbreviations:} NS, neutron star; GWs,
gravitational waves}
%---------------------------------------------------------------------

%---------------------------------------------------------------------
\section{Introduction}\label{sec:intro}
%---------------------------------------------------------------------

A key challenge of modern astrophysics is to obtain the information on the
internal structures of neutron stars (NSs). The precession of deformed NSs
can modulate the precise pulsar timing signals
\citep{Link:2001zr,Jones:2000ud} and produce continuous gravitational waves
\citep[GWs;][]{Zimmermann:1980ba,Jones:2001yg}, which provides a means to
probe the structures of NSs.

In this short contribution, we describe the dynamics of freely-precessing
triaxial NSs in Sec.~\ref{sec:free}. The continuous GWs, electromagnetic
timing residuals, and pulse-width modulations are studied in
Secs.~\ref{sec:gwp} and~\ref{sec:psr}. As the electromagnetic torque acts
on the pulsars in reality, we discuss the vacuum electromagnetic torque and
give an illustrative example of the motions in Sec.~\ref{sec:forced}.
Finally, we summarize our work in Sec.~\ref{sec:disc}.

%---------------------------------------------------------------------
\section{Free precession of triaxial NSs}\label{sec:free}
%---------------------------------------------------------------------

We treat NSs as triaxial rigid bodies and ignore the complex internal
dissipation that may exist. In the body frame corotating with the NS, the
equation of motion takes the following form \citep{landau1960course}
%--
\begin{equation}
\label{eqn:euler}
    \dot {\boldsymbol{L}}+\boldsymbol{\omega} \times \boldsymbol{L}=0 \,,
\end{equation}
%--
where the dot denotes the derivative with respect to time $t$,
$\boldsymbol{\omega}$ is the angular velocity, and
$\boldsymbol{L}=\boldsymbol{I\cdot\omega}$ is the angular momentum with
$\boldsymbol{I}$ represents the moment of inertia tensor. Let
$\boldsymbol{\widehat{e}}_{1}$, $\boldsymbol{\widehat{e}}_{2}$, and
$\boldsymbol{\widehat{e}}_{3}$ denote the three unit eigenvectors along
principal axes of $\boldsymbol{I}$ with corresponding eigenvalues $I_{1}$,
$I_{2}$, and $I_{3}$. Then angular velocity is expressed as
$\boldsymbol{\omega}=\omega_{1}\boldsymbol{\widehat{e}}_{1}+\omega_{2}\boldsymbol{\widehat{e}}_{2}+
\omega_{3}\boldsymbol{\widehat{e}}_{3}$. For simplicity, at $t=0$ we set
$\omega_1=a$, $\omega_2=0$, $\omega_3=b$, and the magnitude of the angular
velocity $\omega=\omega_{0}$.

To describe the precession of NSs, we define 
%--
\begin{equation}
\label{eqn:param}
    \epsilon \equiv \frac{I_{3}-I_{1}}{I_{1}}\,, \quad \delta \equiv \frac{I_{2}-I_{1}}{I_{3}-I_{2}}\,, 
   \quad \theta_{0} \equiv \arctan \frac{I_{1} a}{I_{3} b}\, ,
\end{equation}
%--
where $\epsilon$ is the oblateness, $\delta$ is the non-axisymmetry, and
$\theta_{0}$ is the initial wobble angle between $\boldsymbol{L}$ and
$\boldsymbol{\widehat{e}}_{3}$. To give an estimation of $\epsilon$, we
consider two main causes of deformations: elasticity in the crust and the
internal magnetic field. A small fraction of rotational bulge can misalign
with the instantaneous rotation axis due to the elastic stress in the
crystallized crust of NSs \citep{Cutler:2002np,Melatos:2000qt}. For a NS
with mass $M$, radius $R$ and a constant shear modulus $\mu$ in the crust,
the oblateness from elastic deformation is
\citep{1971AnPhy..66..816B,Gao:2020zcd}
%--
\begin{equation}
    \epsilon_{\rm ela} \simeq 4.9 \times 10^{-8}\left(\frac{\omega_{0}}{2\pi \times100 \,
    \mathrm{Hz}}\right)^{2}\mu_{30}\,R_{6}^{7} M_{1.4}^{-3}\,,
\end{equation}
%--
where $M_{1.4}$, $R_{6}$, and $\mu_{30}$ represent $M/(1.4M_{\odot})$,
$R/(10^{6}\rm cm)$, and $\mu/(10^{30}\,\rm{erg\, cm^{-3}})$ respectively.
While the oblateness caused by internal magnetic field can be roughly
estimated as \citep{Zanazzi:2015ida,Lasky:2013bpa}
%--
\begin{equation}
    \epsilon_{\rm{mag}} \simeq \frac{B^{2}R^{4}}{GM^{2}}=2\times 10^{-12} B_{12}^{2}R_{6}^{4}M_{1.4}^{-2}\,,
\end{equation}
%--
where $B$ is the internal magnetic field and $B_{12}$ is $B/(10^{12}
\,\rm{G})$. In general, the combination of the elastic field in the crust
and the internal magnetic field deform the NS into a triaxial shape and
the non-axisymmetry can be any positive value. The biaxial case is a good
approximation only if one of the deformation causes can be ignored compared
to the other ($\epsilon_{\rm ela}\ll \epsilon_{\rm mag}$ or $\epsilon_{\rm
ela}\ll \epsilon_{\rm mag}$), and the deformation is symmetric about a
specific axis instead of the rotational axis \citep{Melatos:2000qt}.

Eq.~(\ref{eqn:euler}) can be solved analytically. \Blue{With a more
convenient dimensionless parameterization than that in \citet{Gao:2020zcd},
we obtain the time evolution of $u_{i} \equiv \omega_{i}/\omega_{0}$ ($i=1,
2, 3$) in terms of elliptic functions $\mathtt{cn}$, $\mathtt{sn}$, and
$\mathtt{dn}$ \citep{landau1960course,Zimmermann:1980ba}. This new
parameterization is presumed to have a more stable numerical behavior, in
particular when a torque term is added (see Sec.~\ref{sec:forced}).} When
$L^{2}>2EI_{2}$, the solution takes the following form
%--
\begin{align}
    &u_{1}(t)= u_{10}\, \mathtt{cn} (\omega_{\rm p}\,t, m) \label{eqn:omega1}\,,\\
    &u_{2}(t)=u_{10} \left[\frac{(1+\delta)^2}{1+\delta+\epsilon \delta}\right]^{1/2}\mathtt{sn} (\omega_{\rm p}\,t,
    m)\,,\\
    &u_{3}(t)=u_{30}\,\mathtt{dn}(\omega_{\rm p}\,t, m)\,,
\end{align}
%--
for $u_{1}=u_{10} \equiv a/\omega_0$, $u_{2}=0$, and $u_{3}=u_{30} \equiv
b/\omega_0$ at $t=0$, where the parameters $\omega_{\rm p}$ and $m$ are
%--
\begin{align}
    & \omega_{\rm p}= u_{30}\,\omega_{0}\,\epsilon \left({1+\delta+\delta \epsilon}\right)^{-1/2} \,,\\     
    & m= \delta(1+\epsilon)\tan^{2}\theta_{0}\,.
\end{align}
%--
The dimensionless variables $u_{i}$ are periodic with a period 
%--
\begin{equation}
    T=\frac{4K(m)(1+\delta+\delta\epsilon)^{1/2}}{u_{30}\,\omega_{0}\,\epsilon}\,,
\end{equation}
%--
where $K(m)$ is the complete elliptic integral of the first kind
\citep{landau1960course}. In the biaxial case ($\delta=0$ or $\infty$), the
elliptic integral $K(m)$ becomes $\pi/2$, leading to $T=2\pi/\omega_{\rm
p}$.

The motion of the NS in the inertial frame can be described by three Euler
angles: $\phi$, $\theta$, and $\psi$. We take the unit basis vectors of the
coordinate system in the inertial frame as $\boldsymbol{\widehat{e}}_{\rm
X}$, $\boldsymbol{\widehat{e}}_{\rm Y}$, and $\boldsymbol{\widehat{e}}_{\rm
Z}$. We let $\boldsymbol{\widehat{e}}_{\rm Z}$ parallel to
$\boldsymbol{L}$ and define
$\boldsymbol{\widehat{N}}=\boldsymbol{\widehat{e}}_{\rm Z}\times \boldsymbol{\widehat{e}}_{\rm 3}
$. Then, the Euler angles satisfy
%--
\begin{equation}
    \cos \phi=\boldsymbol{\widehat{e}}_{\rm X} \cdot \boldsymbol{\widehat{N}}\,, 
    \quad \cos \theta = \boldsymbol{\widehat{e}}_{\rm 3} \cdot \boldsymbol{\widehat{e}}_{\rm Z}\,, 
    \quad  \cos \psi=\boldsymbol{\widehat{e}}_{1} \cdot\boldsymbol{\widehat{N}}\,.
\end{equation}
%--
The angles $\theta$ and $\psi$ are both periodic with period $T$
\citep{landau1960course}
%--
\begin{align}
    &\cos \theta=\cos \theta_{0} \,\mathtt{dn}(\omega_{\rm p} t, m) \,,\\
    &\tan \psi=\left[\frac{1}{1+\delta+\delta \epsilon}\right]^{1 / 2} 
    \frac{\mathtt{cn}(\omega_{\rm p} t, m)}{\mathtt{sn}(\omega_{\rm p}t, m)}\,.
\end{align}
%--
The angle $\phi$ equals to $\phi_{1}+\phi_{2}$, where \citep{landau1960course}
%--
\begin{align}
    \exp \left[2 \mathrm{i} \phi_{1}(t)\right]&=\frac{\vartheta_{4}\left(\frac{2 \pi }{T}t+\mathrm{i} 
    \pi \alpha, q\right)}{\vartheta_{4}\left(\frac{2 \pi }{T}t-\mathrm{i} \pi \alpha, q\right)}\,, \\
\label{eqn:phi2}
    \phi_{2}=\frac{2 \pi }{T_{1}}t&=\left(\frac{(1+\epsilon)u_{30}\,\omega_{0}}{\cos \theta_{0}}+
    \frac{2 \pi \mathrm{i}}{T} \frac{\vartheta_{4}^{\prime}(\mathrm{i} \pi \alpha, q)}{\vartheta_{4}(\mathrm{i} 
    \pi \alpha, q)}\right) t\,.
\end{align}
%--
Here $\vartheta_{4}$ is the fourth Jacobian theta functions with the nome
$q=\exp [-\pi K(1-m) / K(m)]$, and $\alpha$ can be obtained via
$\mathtt{sn}[2 \mathrm{i} \alpha K(m)]=\mathrm{i} \cot \theta_{0}$. Once
the values of $\epsilon$, $\delta$, $\theta_{0}$, and $\omega_{0}$ are
given, one gets the time evolution of the NS at any time from
Eqs.~(\ref{eqn:omega1}--\ref{eqn:phi2}).

%---------------------------------------------------------------------
\section{Continuous GWs }\label{sec:gwp}
%---------------------------------------------------------------------

Free precession can be manifested in continuous GWs. If the precessing
triaxial NS rotates rapidly, the timing varying mass quadrupole generates
continuous GWs lie in the kilohertz (kHz) band, which is to be observed by
ground-based GW detectors like LIGO, Virgo, and KAGRA.

The general GW waveforms for freely precessing triaxial NSs are
\citep{Zimmermann:1980ba,VanDenBroeck:2004wj}
%--
\begin{align}
    &h_{+} =-\frac{G}{rc^{4}} \big[\left( \mathcal{R}_{2 i} \cos \iota +
    \mathcal{R}_{3 i} \sin \iota\right)\left( \mathcal{R}_{2 j}
    \cos \iota + \mathcal{R}_{3 j} \sin \iota\right) \nonumber \\
    &\quad \quad \quad \quad -\mathcal{R}_{1 i} \mathcal{R}_{1
    j}\big] A_{i j} \label{eqn:waveform_plus}\,,\\
    &h_{\times} =-\frac{2G}{rc^{4}}\left( \mathcal{R}_{2 i} \cos \iota +
    \mathcal{R}_{3 i} \sin \iota\right) \mathcal{R}_{1 j}
    A_{i j}\label{eqn:waveform_cross}\,,
\end{align}
%--
where $h_{+}$ and $h_{\times}$ are ``+'' and ``$\times$'' polarized GWs,
$r$ is the luminosity distance to the NS, $\iota$ is the inclination angle
between the line of sight and $\boldsymbol{\widehat{e}}_{\rm Z}$. Here
$\mathcal{R}_{i j}$ is the rotation matrix, which can be represented by
$\phi$, $\theta$, and $\psi$. The tensor $A_{i j}$ is a function of the
moment of inertia tensor $\boldsymbol{I}$ and the angular velocities
$\omega_{i}$ (see Eq.~(21) in \citet{Zimmermann:1980ba}). In the frequency
domain, the emission of GWs mainly occurs at frequencies
%--
\begin{equation}
    f_{\rm r} + (2n+1) f_{\rm p}\,, \quad 2f_{\rm r} + 2n f_{\rm p}\,,
\end{equation}
%--
where $f_{\rm p}=1/T$ is the free precession frequency, $f_{\rm r}$ equals
to $1/T_{1}-1/T$, and $n$ is an integer.

In the case of small oblatenesses, small wobble angles and small
non-axisymmetries, one can use Taylor expansions of $\theta$ and $\delta$
to obtain the waveforms. The first order contributions occurs at $f_{\rm
r}+f_{\rm p}$ and $2f_{\rm r}$ \citep{VanDenBroeck:2004wj}, with amplitudes
%--
\begin{align}
    {A_{\times}^{1}}= & {2\times 10^{-28}}\,\theta_{0} \sin \iota
    \left(\frac{\epsilon}{10^{-8}} \right) \left(
    \frac{f_{\mathrm{r}}}{100\,\mathrm{Hz}} \right)^{2}
    \left(\frac{1\,\mathrm{kpc}}{r}\right) \,,\\
    {A_{\times}^{2}}=& {4\times 10^{-28}}\,\delta \cos\iota \left(
    \frac{\epsilon}{10^{-8}} \right) \left(
    \frac{f_{\mathrm{r}}}{100\,\mathrm{Hz}} \right)^{2} \left(
    \frac{1\,\mathrm{kpc}}{r}\right) \,,
\end{align}
%--
for ``$\times$'' polarized GWs. The amplitudes for ``+'' polarized GWs can
be obtained similarly but with a different dependence on the inclination
angle $\iota$. The second order lines are too weak \citep{Gao:2020zcd} for
current observational interests and we do not discuss them here.

\section{Modulations in pulsars}
\label{sec:psr}

For pulsars, the emission beam will rotates around the principal axis
$\boldsymbol{\widehat{e}}_{3}$ during free precession and one can possibly
observe timing residuals and pulse profile modulations with radio/X-ray
telescopes \citep{Jones:2000ud,Link:2001zr,Ashton:2016vjx, Gao:2020zcd}.

To study the timing residuals, we assume that the emission is along the
magnetic dipole $\boldsymbol{m}$ and one can observe the pulsar signals
once $\boldsymbol{m}$ sweeps through the plane defined by the line of sight
and $\boldsymbol{\widehat{e}}_{\rm Z}$. We define
%--
\begin{equation}
    \cos \chi = \boldsymbol{\widehat{m}} \cdot \boldsymbol{\widehat{e}}_{\rm 3}\,, 
    \quad \cos \Phi=\boldsymbol{\widehat{e}}_{\rm X} \cdot \boldsymbol{\widehat{M}}\,, 
    \quad \cos \Theta = \boldsymbol{\widehat{m}} \cdot \boldsymbol{\widehat{e}}_{\rm Z}\,,
\end{equation}
%--
where $\boldsymbol{\widehat{m}}$ is $\boldsymbol{m}/|\boldsymbol{m}|$,
$\chi$ is the angle between $\boldsymbol{\widehat{m}}$ and $\boldsymbol{\widehat{e}}_{\rm 3}$,
$\boldsymbol{\widehat{M}}=\boldsymbol{\widehat{e}}_{\rm
Z}\times(\boldsymbol{\widehat{m}}\times \boldsymbol{\widehat{e}}_{\rm Z})$
is the unit vector along the projection of $\boldsymbol{m}$ on the
$\boldsymbol{\widehat{e}}_{\rm X}-\boldsymbol{\widehat{e}}_{\rm Y}$ plane,
$\Phi$ and $\Theta$ are the azimuthal and the polar angles of
$\boldsymbol{\widehat{m}}$ in the inertial frame. The azimuthal angle
$\Phi$ can be expressed as \citep{Jones:2000ud}
%--
\begin{equation}
    \Phi=\phi-\frac{\pi}{2}+\arctan \left(\frac{\cos \psi \sin \chi}{\sin
    \theta \cos \chi-\sin \psi \sin \chi \cos \theta}\right)\,,
\end{equation}
%--
and the phase residual due to precession is
\begin{equation}
    \label{eqn:phase_residual}
    \Delta \Phi=\Phi-\langle \Phi\rangle\,,
\end{equation}
which depends on the relative values of the wobble angle $\theta$ and the
angle $\chi$ (see Eqs.~(45) and (48) in \citet{Jones:2000ud}). In
Eq.~(\ref{eqn:phase_residual}), ``$\langle \cdot\rangle$'' means the time
averaged values. The mean spin period of the pulsar is $P_{0}=2\pi/\langle
\dot \Phi\rangle$. Thus, the timing residual of the spin period $P$ is
%--
\begin{align}
    \Delta P = \frac{2\pi}{\dot \Phi} -\frac{2\pi}{\langle \dot \Phi\rangle} 
    \simeq -\frac{P_{0}^{2}}{2\pi}\Delta \dot{\Phi}\,.
\end{align}
%--
To the second order expansions of $\theta_{0}$ and $\delta$ with a small
wobble angle and a small non-axisymmetry, the period residual is
\citep{Gao:2020zcd}
\begin{align}
    \Delta P \approx & \frac{P_{0}^{2}}{2} f_{\mathrm{p}} \theta_{0}(2 \delta+1) 
    \cot \chi \cos \left(2\pi f_{\mathrm{p}} t\right) \nonumber\\
    &+\frac{P_{0}^{2}}{2} f_{\mathrm{p}} \theta_{0}^{2}\left(1+2 \cot ^{2} \chi\right) 
    \cos \left(4\pi f_{\mathrm{p}} t\right)\,.
\end{align}
Note that the period derivative residual $\Delta \dot{P}$ can be obtained
directly by taking the time derivative of $\Delta P$.

The polar angle $\Theta$,
\begin{equation}
    \cos \Theta=\sin \theta \sin \psi \sin \chi+\cos \theta \cos \chi\,,
\end{equation} 
varies as a function of time. The line of sight cuts different region of
the emission cone during free precession and one could observe pulse width
modulations. Adopting a simple cone model, the pulse width $W$ reads \citep{Lorimer:2005misc,Gil:1984ads}
\begin{equation}
    \sin ^{2}\left(\frac{W}{4}\right)=\frac{\sin ^{2}(\rho / 2)-\sin ^{2}(\beta / 2)}
    {\sin (\Theta+\beta) \sin \Theta}\,,
\end{equation}
where $\rho$ is the angular radius of the emission cone, $\beta$ is the
impact parameter corresponding to the closest approach between the magnetic
dipole moment and the line of sight. Detailed results can be found
in \citet{Gao:2020zcd}. Pulse-width modulations can be used to infer the
emission shape and pulsar radiation properties \citep{Link:2001zr}.

%---------------------------------------------------------------------
\section{Forced precession due to an electromagnetic torque}\label{sec:forced}
%---------------------------------------------------------------------

\Blue{In our previous work \citep{Gao:2020zcd}, we assume that the
precession is free, namely without a torque. However, in reality the
existence of electromagnetic torque will modulate the free precession and
affect continuous GWs and pulsar signals. We take a vacuum electromagnetic
torque as an example
\citep{Goldreich:1970ads,Jones:2001yg,Zanazzi:2015ida}}
%--
\begin{equation}
   \boldsymbol{T} =\boldsymbol{T}_{1}+\boldsymbol{T}_{2}=\frac{2 \omega^{2}}{3 c^{3}}(\boldsymbol\omega \times 
   \boldsymbol m) \times \boldsymbol m+\frac{3}{5R c^{2}}(\boldsymbol \omega \cdot \boldsymbol m)
   (\boldsymbol \omega \times \boldsymbol m)\,,
\end{equation}
%--
and ignore the complex magnetospheric processes
\citep{Arzamasskiy:2015lza}. The first term, $\boldsymbol{T}_{1}$, is the
secular spin down torque, which has components both parallel and
perpendicular to the angular momentum $\boldsymbol{L}$. The component
parallel to $\boldsymbol{L}$ accounts for the usual spin down and the
component perpendicular to $\boldsymbol{L}$ is responsible for the change
of the wobble angles \citep{Jones:2001yg}. The second term,
$\boldsymbol{T}_{2}$, is the anomalous torque, which originates from the
moment of inertia of magnetic dipole field
\citep{Melatos:2000qt,Zanazzi:2015ida}. This torque is perpendicular to the
angular velocity $\boldsymbol{\omega}$ and does not decrease the energy or
the angular momentum. However, it changes the spin period and the wobble
angles for precessing NSs \citep{Jones:2001yg}.

\begin{figure}[t]
    \centerline{\includegraphics[height=13pc,width=80mm]{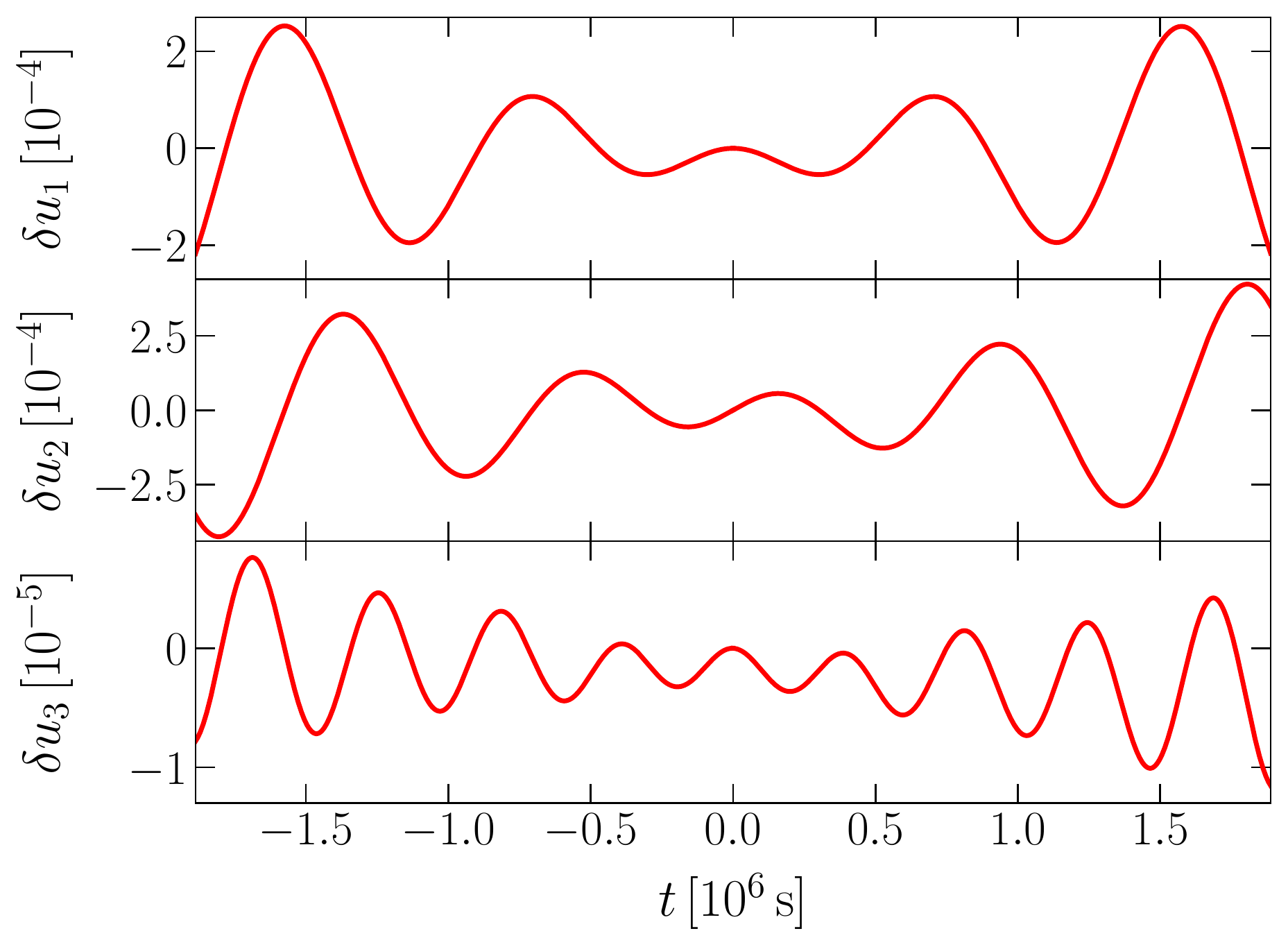}}
    \caption{An illustrative example for the residuals $\delta u_{1}$,
    $\delta u_{2}$, and $\delta u_{3}$ due to electromagnetic torque
    $\boldsymbol{T}$. Here we take $\epsilon=10^{-5}$, $\delta=1$,
    $\theta_{0}=3^{\circ}$, $\omega_{0}=1 \,\rm{rad\,s^{-1}}$, and
    $\chi=89.5^{\circ}$. The time scales $\tau_{\rm c}$ and $\tau_{\rm p}$
    are set to be $10^{12}\,\rm s$ and $1.11\times 10^{8}\,\rm s$.
    \label{fig:forced}}
\end{figure}

The Euler equation for the forced precession with $\boldsymbol{T}$ is 
\citep{Arzamasskiy:2015lza}
%--
\begin{align}
    \dot{\boldsymbol u}+&\frac{1}{ \tau_{\rm p}}\left(\frac{\delta}{1+\delta} \dot{u}_{2} 
    \boldsymbol{\widehat{e}}_{2}+\dot{u}_{3} 
    \boldsymbol{\widehat{e}}_{3}+\frac{\delta}{1+\delta} u_{2}\, \boldsymbol{u} \times 
    \boldsymbol{\widehat{e}}_{2}+u_{3} \boldsymbol{u} 
    \times \boldsymbol{\widehat{e}}_{3}\right) \nonumber\\
   =& \frac{u^{2}}{ \tau_{\rm c}} \left[(\boldsymbol{u}\cdot \boldsymbol{\widehat{m}}) \,
   \boldsymbol{\widehat{m}}-\boldsymbol{u}\right] 
   +\frac{1}{\tau_{a}}\left[(\boldsymbol{u}\cdot \boldsymbol{\widehat{m}}) \,(\boldsymbol{u} 
   \times \boldsymbol{\widehat{m}})\right]\,.
\end{align}
%--
The precession time scale $\tau_{\rm p}$, the secular spin down time scale
$\tau_{\rm c}$ (corresponding to $\boldsymbol{T}_{1}$), and the time scale
$\tau_{\rm a}$ (corresponding to $\boldsymbol{T}_{2}$) are, respectively,
%--
\begin{equation}
    \tau_{\rm p}=1/\epsilon\omega_{0}\,,~~~  \tau_{\rm c}={3 c^{3} I_{1}}/{2 m^{2} \omega_{0}^{2}}
    \,,~~~ \tau_{\rm a}=10 R \omega_{0}\tau_{\rm c}/9c\,,
\end{equation}
%--
where the relation $\tau_{\rm p}\ll \tau_{\rm a}\ll \tau_{\rm c}$ is
usually satisfied \citep{Arzamasskiy:2015lza}. We show an illustrative
example in Fig.~\ref{fig:forced} for forced precession. One notices that
$u_{i}$ oscillates during the forced precession, which is mainly caused by
the torque $\boldsymbol{T}_{2}$. In this small initial wobble angle case,
the angular velocity $\omega \simeq \omega_{0}u_{3}$ and $u_{3}$ decreases
on a longer time scale caused by the spin down torque $\boldsymbol{T}_{1}$.
Because of the relation $\tau_{\rm p}\ll \tau_{\rm a}\ll \tau_{\rm c}$, one
can actually treat torques as perturbation on the free precession in
multi-timescale analysis \citep{Arzamasskiy:2015lza,Link:2001zr}.

The continuous GWs and pulsar signals from forced-precessing triaxial NSs
have some new features and we plan to present them in detail in future
studies.

%--
%--

%---------------------------------------------------------------------
\section{Summary}\label{sec:disc}
%---------------------------------------------------------------------

We gave the analytical solutions of freely-precessing triaxial NSs and
studied their characteristics in continuous GWs and pulsar signals. These
results are ready to be used for future searches of precession. We also
discussed the effects of the electromagnetic torque on the motions of
precession, which deserve more studies in the future. From observations,
precession will give us important information on the equation of state of
NSs and related astrophysical properties~\citep{Gao:2020zcd}.

%\backmatter

%---------------------------------------------------------------------
\section*{Acknowledgments}
%---------------------------------------------------------------------

We are grateful to Rui Xu, Ling Sun, Chang Liu, and Ren-Xin Xu for discussions.
This work was supported by the \fundingAgency{National Natural Science
Foundation of China} under Grant Nos. \fundingNumber{11975027, 11991053,
11721303}, the Young Elite Scientists Sponsorship Program by the
\fundingAgency{China Association for Science and Technology} under the
Grant No. \fundingNumber{2018QNRC001}, and the \fundingAgency{Max Planck
Society} through the \fundingNumber{Max Planck Partner Group}. It was
partially supported by the Strategic Priority Research Program of the
\fundingAgency{Chinese Academy of Sciences} under the Grant No.
\fundingNumber{XDB23010200}, and the High-performance Computing Platform of
Peking University.

%---------------------------------------------------------------------
\subsection*{Conflict of interest}
%---------------------------------------------------------------------

The author declares no potential conflict of interests.

%% \nocite{*}
\bibliography{refs}%

% \section*{Author Biography}
% (if applicable)

\end{document}